\documentclass[letter]{aa}  
\PassOptionsToPackage{dvipsnames}{xcolor}
\usepackage{graphicx}
\usepackage{txfonts}
\usepackage{amsfonts}
\usepackage{physics}
\usepackage{lmodern}
\usepackage{microtype}
\usepackage{booktabs}
\usepackage{bm}
\usepackage{listings}
\usepackage[hidelinks]{hyperref}
\usepackage{subfigure}
\usepackage{mathtools}
\usepackage{natbib}
\usepackage{comment}
\usepackage{orcidlink}

\newcommand{\chieff}{\ensuremath{\chi_\mathrm{eff}}}
\newcommand{\msun}{\mathrm{M}_\odot}
\newcommand{\mcrit}{\mathrm{M}_\mathrm{crit}}

\bibpunct{(}{)}{;}{a}{}{,} 

\begin{document}

\titlerunning{A subpopulation of aligned massive BBHs observed via GWs}
\authorrunning{S.~Rinaldi et al.}
\title{When the black holes align: a subpopulation of aligned massive binary black holes observed via gravitational waves}
\author{
    Stefano~{Rinaldi}\,\inst{1}\orcidlink{0000-0001-5799-4155}\,\thanks{E-mail: stefano.rinaldi@uni-heidelberg.de}
    \and
    Charmaine~{Wong}\,\inst{2}\orcidlink{0009-0007-9367-4205}\thanks{E-mail: charmainewong@link.cuhk.edu.hk}
    \and
    M. Paola {Vaccaro}\,\inst{1}\orcidlink{0000-0003-3776-9246}\thanks{E-mail: mariapaolavaccaro@gmail.com}
    \and
    Elvir~{Mislimi}\,\inst{1}\orcidlink{0009-0005-3538-2234}
    \and
    Otto~A.~{Hannuksela}\,\inst{2}\orcidlink{0000-0002-3887-7137}
    \and
    Samson~H.~W.~{Leong}\,\inst{2}\orcidlink{0000-0003-0470-282X}
    \and
    Michela~{Mapelli}\,\inst{1}\orcidlink{0000-0001-8799-2548}
    }
\institute{
    Institut für Theoretische Astrophysik, ZAH, Universität Heidelberg, Albert-Ueberle-Str.~2, 69120 Heidelberg, Germany
    \and
    Department of Physics, The Chinese University of Hong Kong (CUHK), Shatin, New Territories, Hong Kong
    }
\date{Received \today; accepted XXX}
\abstract
{}
{In this work, we investigate the features present in the joint primary mass and effective spin distribution of binary black holes without relying on specific modelling assumptions.}
{We make use of non-parametric methods, flexible models capable of approximating arbitrary probability densities with minimal mathematical assumptions, applying them to the newly released GWTC-5.0.}
{Our analysis supports, albeit with large uncertainties, the presence of at least two separate sub-population of binary black holes showing different effective spin distributions: one of them with large primary mass ($m_1\gtrsim{}50$ M$_\odot$) and a preference for positive $\chieff$ values points towards systems formed in a non-spherically-symmetric, dynamical environment.}
{}


\maketitle
\section{Introduction}
With more than 250 bona-fide gravitational-wave (GW) signals \citep{gwtc5:2026} detected by the LIGO-Virgo-KAGRA collaboration in the ten years since GW150914 \citep{gw150914:2015}, we are now in an era in which our understanding of the origin of compact binary systems is spearheaded by detailed population studies. 
We can now look for correlations among binary black hole (BBH) parameters at the population level. For instance, \cite{callister:2021} report an anti-correlation between  the mass ratio $q$ and the effective spin parameter $\chieff=(\mathrm{M}_1\,{}\vec{\chi}_1+\mathrm{M}_2\,{}\vec{\chi}_2)/(\mathrm{M}_1+\mathrm{M}_2)\cdot{}\vec{L}/L$, where $\mathrm{M}_1,\,{}\mathrm{M}_2,\,{}\vec{\chi}_1,\,{}\vec{\chi}_2$ are the masses and spin vectors of the two black holes and $\vec{L}$ is the orbital angular momentum. \cite{biscoveanu:2022} report a broadening of the spin distribution with redshift. More recently, \cite{tong:2026}, \cite{antonini:2025:gwtc4}, \cite{flanagan:2026}, and \citet{padhyegurjar:2026} independently report that black holes in the pair-instability mass gap tend to spin faster than lower mass black holes. These correlations, summarised in \citet{biscoveanu:2026}, are a powerful tool to infer the formation channels of BBHs \citep[e.g.][]{gerosa:2021,mapelli:2022,vaccaro:2024,galaudage:2026,wang:2026}. 

Multivariate non-parametric methods -- flexible models that can approximate arbitrary probability densities in multiple dimensions \citep[e.g.][]{rinaldi:2022:hdpgmm,heinzel:2024,tenorio:2025} -- can reveal the presence of these correlations in an agnostic fashion without modelling them a priori.
In this Letter, we investigate the joint $\mathrm{M}_1-\chieff$ distribution using a non-parametric model and report the presence of a sub-population of massive BBHs with spins that are not isotropically distributed, hint of a non-spherically-symmetric dynamical formation scenario. \vspace{-0.1cm}

\section{Joint $\mathrm{M}_1-\chieff$ distribution}\label{sec:results}
\begin{figure}
    \centering
    \subfigure{\includegraphics[width=0.8\columnwidth]{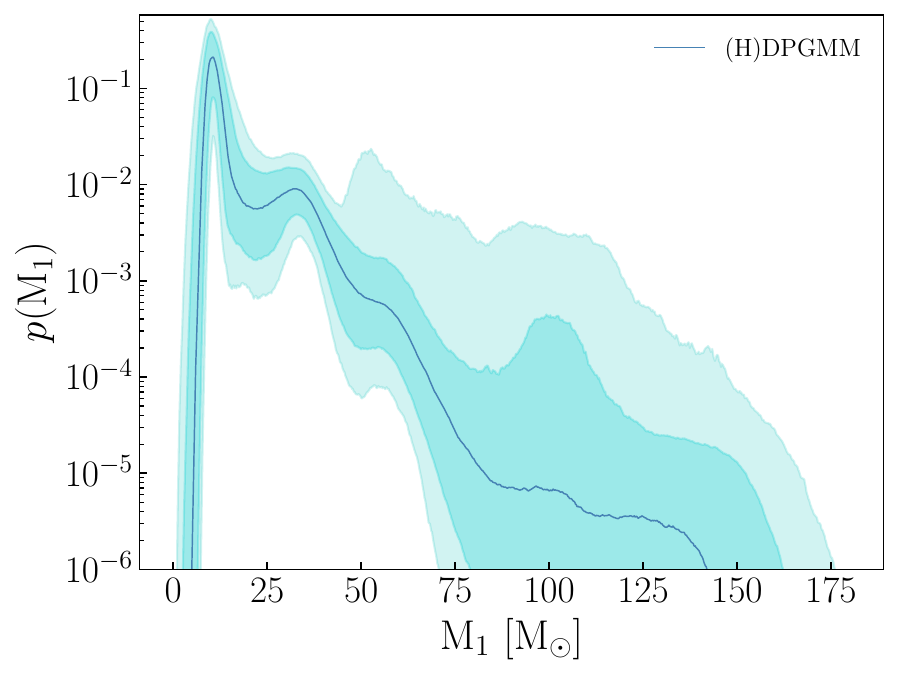}\label{fig:m1_marg}
    }
    \subfigure{\includegraphics[width=0.8\columnwidth]{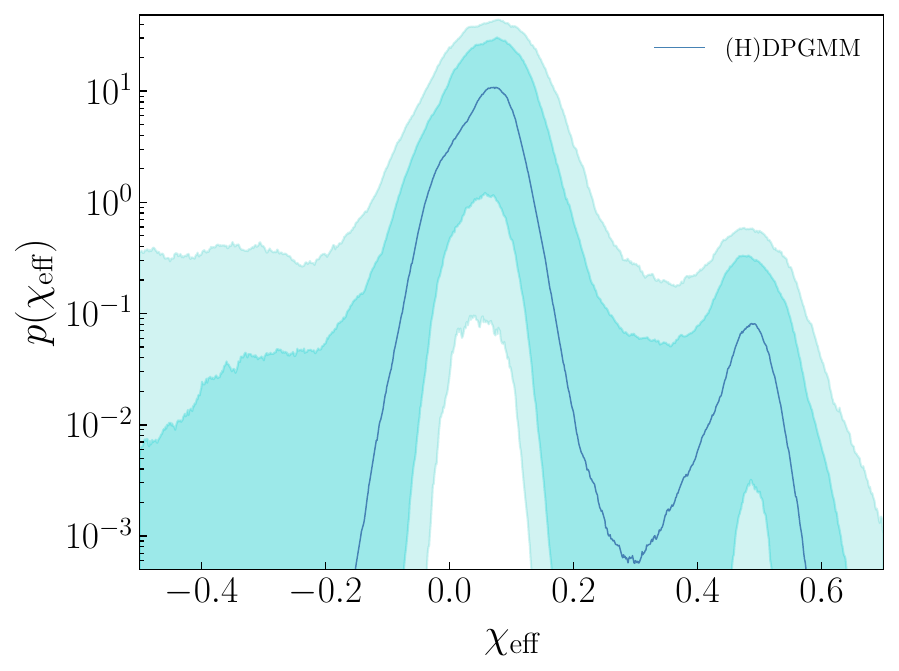}\label{fig:chieff_marg}
    }
    \caption{Marginal primary mass (top) and effective spin (bottom) distributions reconstructed using (H)DPGMM. The solid line marks the median distribution, whereas the shaded areas correspond to the 68\% and 90\% credible regions.}
    \label{fig:marginals}
\end{figure}
Here, we present the results of our inference of the joint $\mathrm{M}_1-\chieff$ distribution, with details on the models and data used given in Appendix~\ref{app:methods}.
Figure~\ref{fig:marginals} reports the inferred non-parametric marginal distributions for $\mathrm{M}_1$ and $\chieff$. The marginal primary mass distribution and the parameters of the redshift ($\kappa = 2.8^{+0.5}_{-0.5}$) and mass ratio ($\beta = 0.6^{+0.7}_{-0.6}$) distributions are in agreement with the findings reported by \citet{gwtc5pop:2026}. 
The main focus of this work is on the inferred joint $\mathrm{M}_1-\chieff$ distribution, reported in Figure~\ref{fig:slices} as the conditional distribution $p(\chieff|\mathrm{M}_1)$ for different values of the primary mass. In it, we identify two main features: a drift from $\chieff \sim 0.1$ to $\chieff \sim 0$ in the low-mass regime and the presence of a second subpopulation of preferentially aligned binaries -- $\chieff \sim 0.4$ -- for primary masses $\mathrm{M}_1 > 40\,\msun$.
A third, smaller feature is visible at $20\,\msun$, $\chieff\sim 0.5$: this stand-alone peak is composed by GW241011\_233834 and GW241113\_163507 and, possibly, by GW231118\_005626. At this stage, however, it is not possible to assess whether these objects belong to a separated subpopulation or if we are in presence of a statistical fluctuation (see Appendix~\ref{app:figures} for further details).

\vspace{-0.2cm}
\subsection{$\mathrm{M}_1 < 50\,\msun$: slow rotation or precession?}
We find that the effective spin distribution of relatively low-mass BBHs strongly prefers slowly spinning (or precessing) systems. In correspondence of the main peak at $10\,\msun$, the $\chieff$ distribution is centred at 0.1 and then smoothly drifts towards vanishing effective spins with increasingly higher masses, up to $\sim 33\,\msun$. The presence of a slowly rotating, aligned subpopulation at low masses is consistent with the hypothesis that the $10 \,\msun$~peak in the primary mass distribution is generated by the isolated evolution channel \citep[e.g.][]{bavera:2020, vanSon:2022}.
A similar trend is reported in \citet{cheng:2026} and in the non-parametric findings of~\citet{gwtc5pop:2026}: both their \textsc{Binned Gaussian Process} and \textsc{PixelPop} models, the results of which are reported in their Figure 6, compares the effective spin distribution of the binaries with primary mass in the $[8,15]\, \msun$ interval -- encompassing the $\mathord{\sim} 10\ \msun$ peak -- with the binaries outside this range. The $10\,\msun$ peak in the mass spectrum is associated with an effective spin distribution supporting preferentially positive values of $\chieff$, whereas the $33\ \msun$ peak corresponds to a more isotropic distribution centred around $\chieff\sim 0$. Our findings suggest, however, that the transition between these two regimes is continuous and smooth.

\subsection{$\mathrm{M}_1 > 50\,\msun$: aligned systems}\label{sec:aligned}
\begin{figure}
    \centering
    \includegraphics[width=0.85\linewidth]{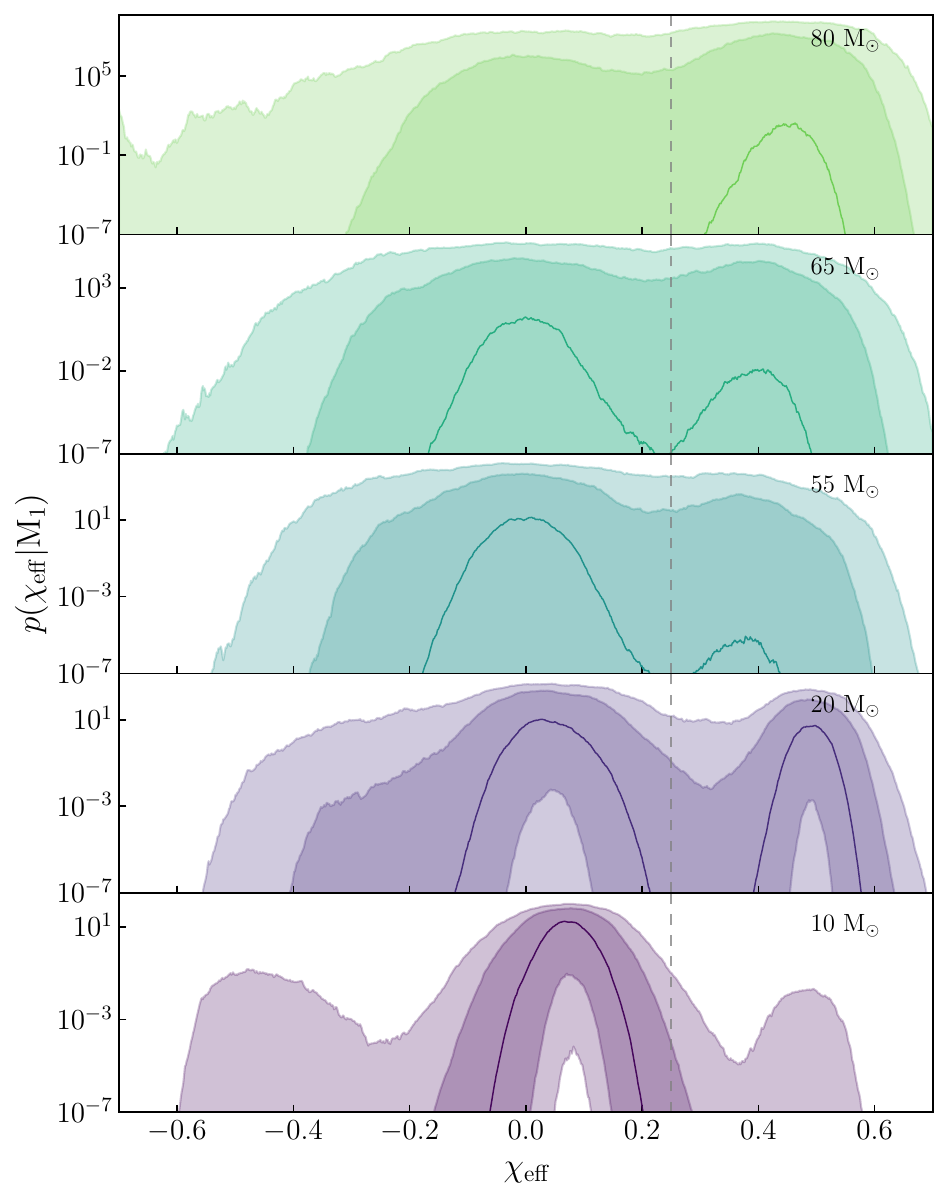}
    \caption{Reconstructed $\chieff$ distribution conditioned on different masses. The vertical dashed line at $\chieff = 0.25$ marks the boundary between the two subpopulations.}
    \label{fig:slices}
\end{figure}

The second feature of the mass-spin distribution is the emergence of a positively aligned subpopulation, characterized by $\chieff>0.25$, which appears predominantly for primary masses above $50\ \msun$.
This subpopulation, as highlighted in Figure~\ref{fig:slices}, does not appear to branch out of the main population discussed in the previous paragraph, but rather to emerge independently above a certain critical mass $\mcrit$, suggesting a separate origin for these aligned systems and therefore at least two formation channels in the observed BBH population. Trends analogous to the ones described here are reported by \citet{antonini:2025:gwtc3} and \citet{antonini:2025:gwtc4}, who, using a parametrised model, infer a transition from a Gaussian distribution of $\chieff$ for light systems to a uniform distribution above 45 $\msun$, and \citet{plunkett:2026} (all using GWTC-4.0) and in \citet{alvarez:2026} and \citet{flanagan:2026} (both with GWTC-5.0). 

Broadly defining the aligned subpopulation as having primary mass larger than $50\,\msun$ and effective spin larger than 0.25, we find that this feature accounts for $0.09^{+0.13}_{-0.06}\%$ of the total BBH population. To obtain more quantitative constraints on this subpopulation, we performed a second analysis using a parametrised model for the effective spin, a weighted superposition of a skewed Gaussian distribution \citep{banagiri:2025:skewed} and a symmetric Gaussian distribution. To account for the fact that the second subpopulation is only present above a certain mass $\mathrm{M_{crit}}$, we include also an activation function $\mathcal{S}(\mathrm{M}_1|\mathrm{M_{crit}},\alpha)$,
\begin{multline}\label{eq:par_chi_model}
    p(\chieff|\mathrm{M}_1) \propto \qty(1-w\,\mathcal{S}(\mathrm{M}_1|\mathrm{M_{crit}},\alpha)) \,{}\mathcal{SN}\qty(\chieff|\mu_1,\sigma_1,\varepsilon) \\ + w\,{}\mathcal{S}(\mathrm{M}_1|\mathrm{M_{crit}},\alpha) \,{}\mathcal{N}\qty(\chieff|\mu_2,\sigma_2)\,,
\end{multline}
whereas the primary mass distribution is still modelled using a non-parametric approach. The functional form for this model is given, together with the full posterior distribution, in Appendix~\ref{app:parmodel}. In order to prevent contamination from the aligned subpopulation at $20\,\msun$, we excluded from this last analysis the three events mentioned at the beginning of this Section.
We find that the second subpopulation, centred at $\mu_2 = 0.47^{+0.06}_{-0.09}$ with $\sigma_2 < 0.15$, starts appearing at $48^{+8}_{-9}\,\msun$ ($\mathrm{M_{crit}}-2\alpha$, see Eq.~\eqref{eq:activation}) in agreement with the findings of analogous works using GWTC-4.0 \citep{antonini:2025:gwtc3,banagiri:2025,antonini:2025:gwtc4,hussain:2026,tong:2026,plunkett:2026}. The skewness parameter of the skewed Gaussian distribution vanishes ($\varepsilon = 0.16^{+0.24}_{-0.26}$): this is in contrast with the model-dependent skewed distribution reported in \citet{gwtc5pop:2026}, suggesting that their observed skewness is likely due to the presence of this subpopulation \citep[already in][]{plunkett:2026,alvarez:2026}.
To quantitatively compare the model in Eq.~\eqref{eq:par_chi_model} with a second one that uses only one skewed Gaussian distribution, we computed the likelihood ratio $\mathcal{L}$ between the two models. We found that $\log_{10}\mathcal{L} = 7.2 \pm 2.6$, favouring the model with two effective spin subpopulations. This quantity, however, does not account for the different sizes of the prior volumes: accounting for the prior ratio, $\log_{10}\mathcal{P} = 3.5$, rescales the likelihood ratio to a more conservative $\log_{10}\mathcal{L} = 3.8 \pm 2.6$\footnote{It would be more appropriate to use the Bayes' factor for model selection: however, due to the size and degeneracy of the parameter space of the non-parametric model used for the primary mass, the evidence computation necessary for the Bayes factor is, to the best of our knowledge, not possible.}.

\subsection{$\chieff<0.25$: pileup or uniform?}
Our non-parametric reconstruction shows a preference for two independent sub-populations at high masses.
However, due to the short chirp duration of high-mass BBH mergers, the amount of information carried by individual GW events on their effective spin is limited: the precise shape of this subpopulation is hence difficult to determine using non-parametric methods alone, as shown in Figure~\ref{fig:slices}. Several studies \citep{antonini:2025:gwtc3,antonini:2025:gwtc4,flanagan:2026,ray:2026,tong:2026,cheng:2026} investigate and report the presence of a uniform (or broad, in the case of \citealt{cheng:2026}) component in the $\chieff$ distribution associated with massive binaries. The uniform distribution reported in these works encompasses the effective spin range between around -0.5 and 0.5, with extremal values inferred from the data.
In the high-mass regime, the credible areas our inferred distribution cover the same interval, not ruling out the possibility of a uniform distribution explaining both the features we report here.
Our findings are hence not in tension with the papers cited above: these works identify the two subpopulations reported here with a single uniform distribution, and the often almost unconstrained $\chi_\mathrm{eff,min}$ reflects in the large $90\%$ credible region we obtain for negative $\chieff$ values. These uncertainty regions encompass effective spins down to -0.5, in agreement with the presence of counter-rotating BBHs in the catalogue.

\citet{antonini:2025:gwtc4} make use of a non-parametric method based on the Gaussian process \citep{callister:2024} to model the effective spin distribution at high masses. Their method, being inherently one-dimensional, reconstructs the integrated $\chieff$ population above $40\,\msun$ and therefore is not able to recover the specific mass-dependent trend we report here. Nonetheless, both methods support the claim that massive BBHs prefer positive effective spin. 
Using a different multivariate non-parametric method, \citet{alvarez:2026} explore the correlated $\mathrm{M}_1-\chieff$ distribution, finding evidence for the same aligned subpopulation reported in this paper as well as supporting a broad but uncertain distribution for $\chieff$ at high masses.

Moreover, several works \citep[e.g.][]{ray:2026,cheng:2026} show evidence for a transition in the mass ratio $q$ regime as a function of primary mass. Due to the intrinsic correlation between $q$ and $\chieff$, it is possible for the mass-dependent $\chieff$ distribution discussed above to be induced by the correlation between primary mass and mass ratio. The amount of information carried by GWs about $q$ is however limited \citep{rinaldi:2025:features}, and the assumed population model plays a significant role in the findings. In summary, we believe that at the current stage the large uncertainties associated with the $\chieff$ of BBHs limit us to simple hints on the shape of the effective spin distribution for massive binaries based on the data alone.

\vspace{-0.2cm}\section{Astrophysical implications}\label{sec:astro_implications}

We have reported a feature in the joint distribution of primary mass and effective spin ($\mathrm{M}_1-\chieff$). We highlight the existence of a subpopulation with $\mathrm{M}_1 \geq 50\,\msun$ and $\chieff\geq 0.25$ (Fig.~\ref{fig:slices}) which is not naturally described as a smooth continuation of the main low-mass, low-spin population.
A positive effective spin clearly indicates that at least part of the spin of the component BHs is preferentially aligned with the orbital angular momentum, suggesting a formation mechanism that causes spin-orbit alignment.

One possible explanation is hierarchical growth in AGN discs \citep{wang:2021,li:2025:gwtc3,li:2025:gwtc4}. In this scenario, stellar-origin BHs can be captured by the disc, form binaries, and participate in repeated mergers, producing increasingly massive BHs. The gaseous disc provides a preferred angular-momentum direction, so that this channel can naturally generate a population of BHs with both large spin magnitudes and preferentially positive $\chieff$, leading to a clear $\mathrm{M}_1-\chieff$ correlation \citep{vaccaro:2024, vaccaro:2026}. In \autoref{app:AGNs}, we compare the inferred high-mass aligned subpopulation with synthetic catalogues of BBH mergers in AGN discs. 

Hierarchical mergers in dense stellar clusters can also produce massive BHs with $\mathrm{M}_1>40 \, \msun$. We generally expect spin orientations to be close to isotropic in these channels, which would lead to a $\chieff$ distribution symmetrical around zero, but this expectation is strongly model-dependent \citep[e.g.][]{antonini:2020, mapelli:2022}. In particular, some degree of alignment could arise if residual gas is present  \citep{kirouglu:2025}. Therefore, while the positive-$\chieff$ preference disfavours a fully isotropic hierarchical-merger population as the sole explanation, it does not exclude cluster-assisted hierarchical growth altogether. 

Isolated-binary evolution might also allow the formation of this subpopulation under specific conditions. For instance, mass ratio reversal and tidal spin-up of the secondary star might lead to a rapidly rotating second-born massive BH,  whose spin is preferentially aligned with the orbital angular momentum \citep{bavera:2021,kapil:2026}. Alternatively, stochastic convective motions in a red-supergiant progenitor may provide sufficient angular momentum to produce a BH with appreciable spin during collapse \citep{antoni:2022}. Finally, 
chemically homogeneous evolution \citep{demink:2016, marchant:2016} is expected to naturally lead to massive BBHs with large and preferentially aligned spins \citep{marchant:2024}. In this channel, efficient internal mixing keeps stars compact, while tidal coupling in close binaries can spin them up and promote spin--orbit alignment. The main open question concerning the isolated binary (or triple) evolution scenario would be whether it can produce primary black holes with mass $m_1>50$ M$_\odot$, possibly above the pair-instability mass gap. Recent results indicate that  some configurations of extreme dredge ups \citep{costa:2021}, chemically homogeneous mass transfer (\citealt[e.g.][]{croon2026,popa2025}), and triple system evolution \citep[e.g.][]{stegmann2025} might indeed produce isolated aligned BBHs with primary mass $>50$ M$_\odot$.
 
In summary, both chemically homogeneous evolution and AGN disk dynamics appear as promising mechanisms to explain this $M_1-\chi_{\rm eff}$ feature, as they simultaneously produce high BH masses, large spin magnitudes, and preferential spin--orbit alignment.

\begin{acknowledgements}
The authors are grateful to Aleksandra~Olejak, Jakob~Stegmann, several LVK collaboration members and the anonymous referee for useful discussions and suggestions.

This work was funded by the Deut\-sche For\-schungs\-ge\-mein\-schaft (DFG, German Research Foundation) – project number 546677095 – and from the German Excellence Strategy via the Heidelberg Cluster of Excellence (EXC 2181 - 390900948) STRUCTURES. 
The authors acknowledge support by the state of Baden-W\"urttemberg through bwHPC, the German Research Foundation (DFG) through grants INST 35/1597-1 FUGG and INST 35/1503-1 FUGG, Research Grants Council of Hong Kong (Project No. CUHK 14304622, 14307923, and 14307724), the start-up grant from the Chinese University of Hong Kong, and the Direct Grant for Research from the Research Committee of The Chinese University of Hong Kong.

This research has made use of data or software obtained from the Gravitational Wave Open Science Center (gwosc.org), a service of the LIGO Scientific Collaboration, the Virgo Collaboration, and KAGRA. This material is based upon work supported by NSF's LIGO Laboratory which is a major facility fully funded by the National Science Foundation, as well as the Science and Technology Facilities Council (STFC) of the United Kingdom, the Max-Planck-Society (MPS), and the State of Niedersachsen/Germany for support of the construction of Advanced LIGO and construction and operation of the GEO600 detector. Additional support for Advanced LIGO was provided by the Australian Research Council. Virgo is funded, through the European Gravitational Observatory (EGO), by the French Centre National de Recherche Scientifique (CNRS), the Italian Istituto Nazionale di Fisica Nucleare (INFN) and the Dutch Nikhef, with contributions by institutions from Belgium, Germany, Greece, Hungary, Ireland, Japan, Monaco, Poland, Portugal, Spain. KAGRA is supported by Ministry of Education, Culture, Sports, Science and Technology (MEXT), Japan Society for the Promotion of Science (JSPS) in Japan; National Research Foundation (NRF) and Ministry of Science and ICT (MSIT) in Korea; Academia Sinica (AS) and National Science and Technology Council (NSTC) in Taiwan.
\end{acknowledgements}
\bibliographystyle{aa}
\bibliography{references.bib}

\begin{appendix}
\section{Models, data, and methods}\label{app:methods}
We include, in our analysis, four binary parameters, namely the primary mass $\mathrm{M}_1$, the mass ratio $q$, the redshift $z$ and the effective spin $\chieff$, marginalising over the others as described in Appendix C.5 of \cite{essick:2025}. Our astrophysical population model is designed to be separable, 
\begin{equation}p(\mathrm{M}_1,q,z,\chieff) = p(\mathrm{M}_1,\chieff)\,p(q)\,p(z)\,.
\end{equation}
The mass ratio follows a power-law distribution, 
\begin{equation}
    p(q|\beta) \propto q^\beta\,, 
\end{equation}
whereas the redshift is proportional to 
\begin{equation}
p(z)\propto (1+z)^{\kappa-1} \frac{\dd V_c}{\dd z}\,.
\end{equation}
For the two-dimensional joint primary mass-effective spin distribution, the main focus of this paper, we make use of (H)DPGMM \citep{rinaldi:2022:hdpgmm}, a hierarchical, non-parametric method based on the multivariate Gaussian mixture model,
\begin{equation}
    p(M_1,\chieff) \simeq \sum_i \,{}w_i \,{}\mathcal{N}(M_1,\chieff|\boldsymbol{\mu}_i,\boldsymbol{\Sigma}_i)\,,
\end{equation}
where $\boldsymbol{\mu}_i$ and $\boldsymbol{\Sigma}_i$ are the mean vector and covariance matrix respectively. The likelihood used for the inference is the scale-free hierarchical likelihood \citep[see e.g.][]{mandel:2019}, and the inference is performed using \textsc{emcee} \citep{emcee} and \textsc{figaro} \citep{rinaldi:2024:figaro} combined in a Gibbs sampling scheme. 
We will make use of the publicly available\footnote{Available at \texttt{\href{https://zenodo.org/records/20348005}{https://zenodo.org/records/20348006}} and \texttt{\href{https://zenodo.org/records/20348006}{https://zenodo.org/records/20348006}}} GW events released by the LVK collaboration as part of the Gravitational Wave Transient Catalogue (GWTC) up to its most recent update, GWTC-5 \citep{gwtc5:2026,opendata:2026}. The events are selected using the same criterion used in \citet{gwtc5pop:2026} for BBH-only analyses (false-alarm rate < 1 yr$^{-1}$, 1\% lower limit on secondary mass larger than 3 $\msun$), leaving us with a grand total of 258 BBH mergers. 

The GW observations and the astrophysical distribution are linked via the hierarchical likelihood. In particular, we make use of the scale-free version of the likelihood \citep[see e.g.][]{mandel:2019}, with the global merger rate $\mathcal{R}_0$ marginalised out using the appropriate prior:
\begin{equation}\label{eq:jointlikelihood}
    p(\Lambda|\mathbf{d}) = \prod_i^N \frac{p(d_i)}{\alpha(\Lambda)} \int \frac{p(\theta_i|d_i)p(\theta_i|\Lambda)}{\pi(\theta_i)}\dd\theta_i\,.
\end{equation}
Here we denoted with $\mathrm{d} = \{d_1,\ldots,d_N\}$ the GW data corresponding to the $N$ available detections, with $\Lambda$ the hyperparameters of the astrophysical models and with $\theta$ the binary parameters. The detectability fraction $\alpha(\Lambda)$ is defined starting from the detection probability $p_\mathrm{det}(\theta)$ as
\begin{equation}
    \alpha(\Lambda) = \int p_\mathrm{det}(\theta)p(\theta|\Lambda)\dd \theta\,,
\end{equation}
and it is estimated making use of a set of simulated signals\footnote{Available at \texttt{\href{https://zenodo.org/records/19500052}{https://zenodo.org/records/19500052}}} in a Monte Carlo sum \citep{essick:2025}:
\begin{equation}
    \alpha(\Lambda) \simeq \frac{1}{N_\mathrm{inj}}\sum_i^{N_\mathrm{found}} \left.\frac{p(\theta_i|\Lambda)}{p_\mathrm{inj}(\theta_i)}\right\vert_{\theta \sim p_\mathrm{inj}(\theta)}\,.
\end{equation}
In this work, we made use of an astrophysical distribution that is a combination of parametric and non-parametric model. Whereas for the inference of a relatively low number of parameters, around $o(30)$, Markov Chain Monte Carlo (MCMC) techniques are effective and well optimised to sample parameter space, non-parametric methods require a much larger number of degenerate parameters,\footnote{The infinite Gaussian mixture model is, in fact, over-complete: therefore, there are infinite optimal combinations of parameters.} often beyond the capabilities of regular MCMC algorithms. For this very reason, non-parametric methods often employ different sampling strategies and rely on ad-hoc numerical algorithms specifically tailored to particular models: this is the case of (H)DPGMM and its associated code \textsc{figaro}, used in this work. The downside of this is that \textsc{figaro} is not able to infer the hyperparameters of a generic parametric model.

To circumvent this issue, we used a Gibbs sampling scheme \citep{geman:1984}, a MCMC algorithm that finds its application when sampling from the joint distribution of a set of parameters is difficult, computationally too expensive or in some cases even impossible, but at the same time sampling from the corresponding conditional distributions can be easily done.
In our specific case, the hyperparameters $\Lambda$ that we want to sample are the spectral indexes $\kappa$ and $\beta$ for the redshift and mass ratio respectively and $\Theta = \{\mathbf{w},\boldsymbol\mu,\boldsymbol\Sigma\}$, the parameters of the Gaussian mixture model, for the joint primary mass and effective spin distribution.
We can then break down the likelihood in Eq.~\eqref{eq:jointlikelihood} in its separate parts,
\begin{multline}
    p(\kappa,\beta,\Theta|\mathbf{d}) = \prod_i^N \frac{p(d_i)}{\alpha(\kappa,\beta,\Theta)} \int p(\theta_i|d_i) \frac{p(q_i, z_i|\kappa,\beta)}{\pi(\theta_i)}\\ \times p(\mathrm{M}_{1,i},\chi_{\mathrm{eff},i}|\Theta) \dd\theta_i\,,
\end{multline}
as well as the Monte Carlo sum used to estimate the detectability fraction:
\begin{equation}
    \alpha(\kappa,\beta,\Theta) \simeq \sum_i^{N_\mathrm{found}} \left.\frac{p(q_i, z_i|\kappa,\beta)p(\mathrm{M}_{1,i},\chi_{\mathrm{eff},i}|\Theta)}{N_\mathrm{inj}\, p_\mathrm{inj}(\theta_i)}\right\vert_{\theta \sim p_\mathrm{inj}(\theta)}\,.
\end{equation}
If we keep either $(\kappa,\beta)$ or $\Theta$ fixed, the corresponding term in the likelihood simply becomes an additional weight to be used in the Monte Carlo estimations of the integrals. Doing so allows us to use dedicated algorithms to draw samples for $\Theta$ and to keep the parameter space explored by conventional MCMC samplers limited in dimensionality.
\section{Additional figures for Section~\ref{sec:results}}\label{app:figures}
\begin{figure}
    \centering
    \includegraphics[width=0.95\columnwidth]{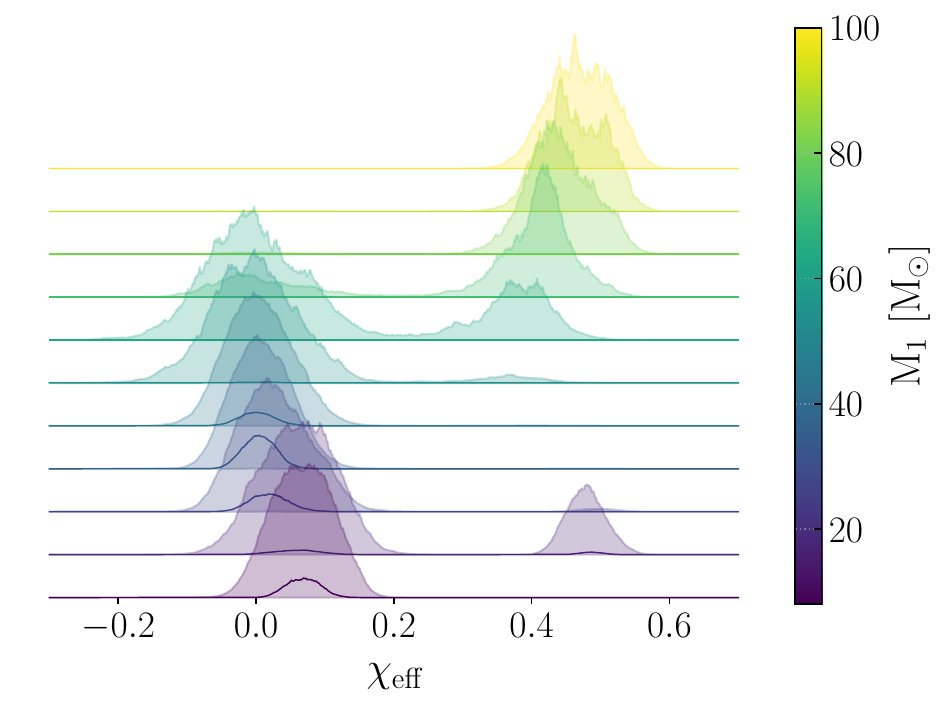}
    \caption{Reconstructed effective spin distribution conditioned on primary mass. The shaded areas mark the $68\%$ credible regions, whereas the solid line marks the median normalised distribution. Slices above $\sim 40\,\msun$ are uncertainty-dominated and therefore the median line is not visible.}
    \label{fig:joyplot}
\end{figure}

\begin{figure*}
    \centering
    \includegraphics[width=1.95\columnwidth]{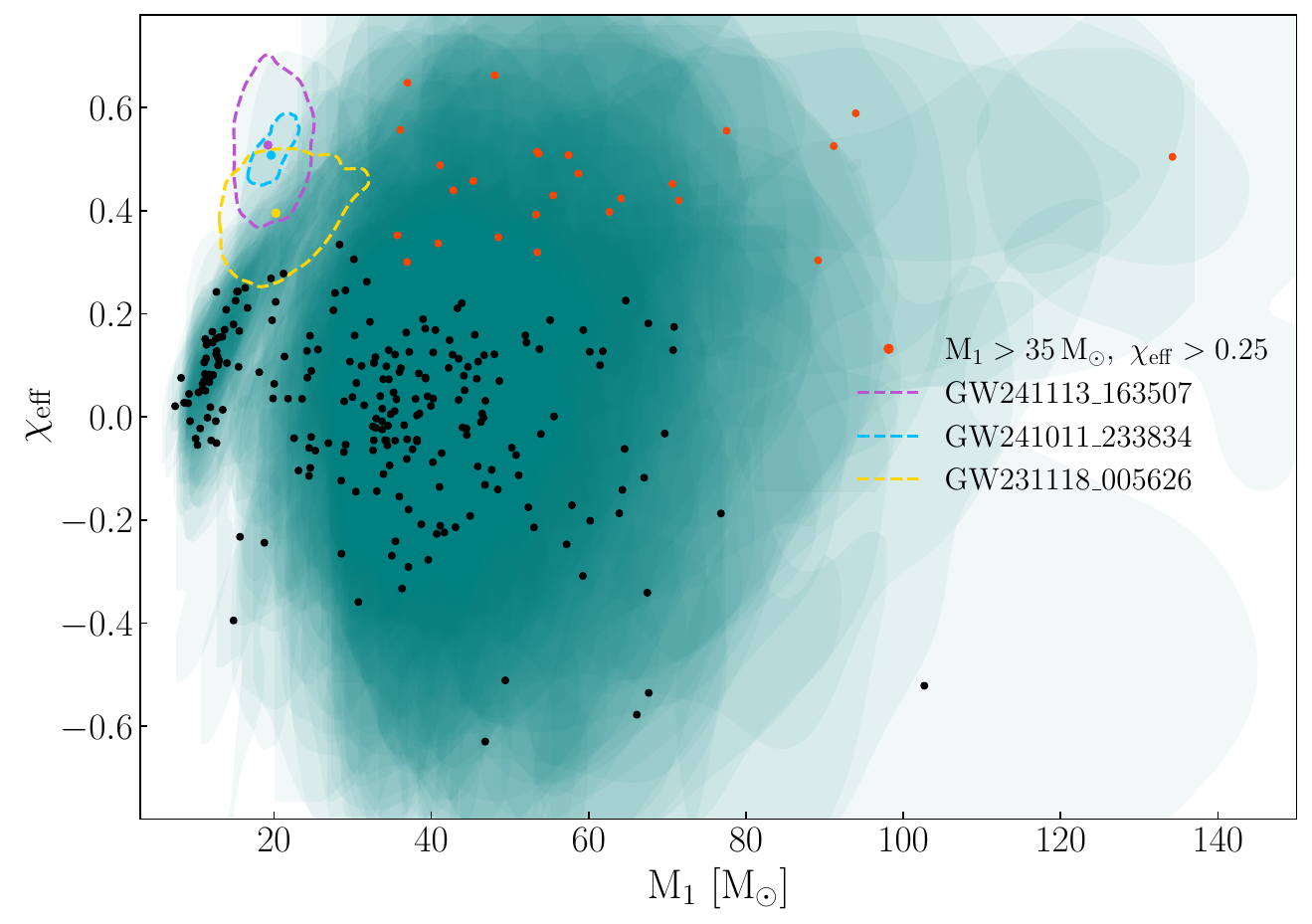}
    \caption{Likelihood distributions for the events in GWTC-5 included in this work reported as median and 90\% credible area. The dashed contours highlight the three events that are likely part of the $20\,\msun$ aligned subpopulation, whereas the red dots mark the events with median $\mathrm{M}_1 > 35\,\msun$ and median $\chieff > 0.25$.}
    \label{fig:likelihoods}
\end{figure*}

In Figure~\ref{fig:joyplot}, we show the mass-dependent $\chieff$ distribution on a linear scale. Figure~\ref{fig:likelihoods} shows the likelihoods of the 258 events included in our analysis, highlighting the events that are likely to be part of the two aligned subpopulations.

\section{Parametrised mass-dependent $\chieff$ distribution}\label{app:parmodel}

\begin{figure*}
    \centering
    \includegraphics[width=1.95\columnwidth]{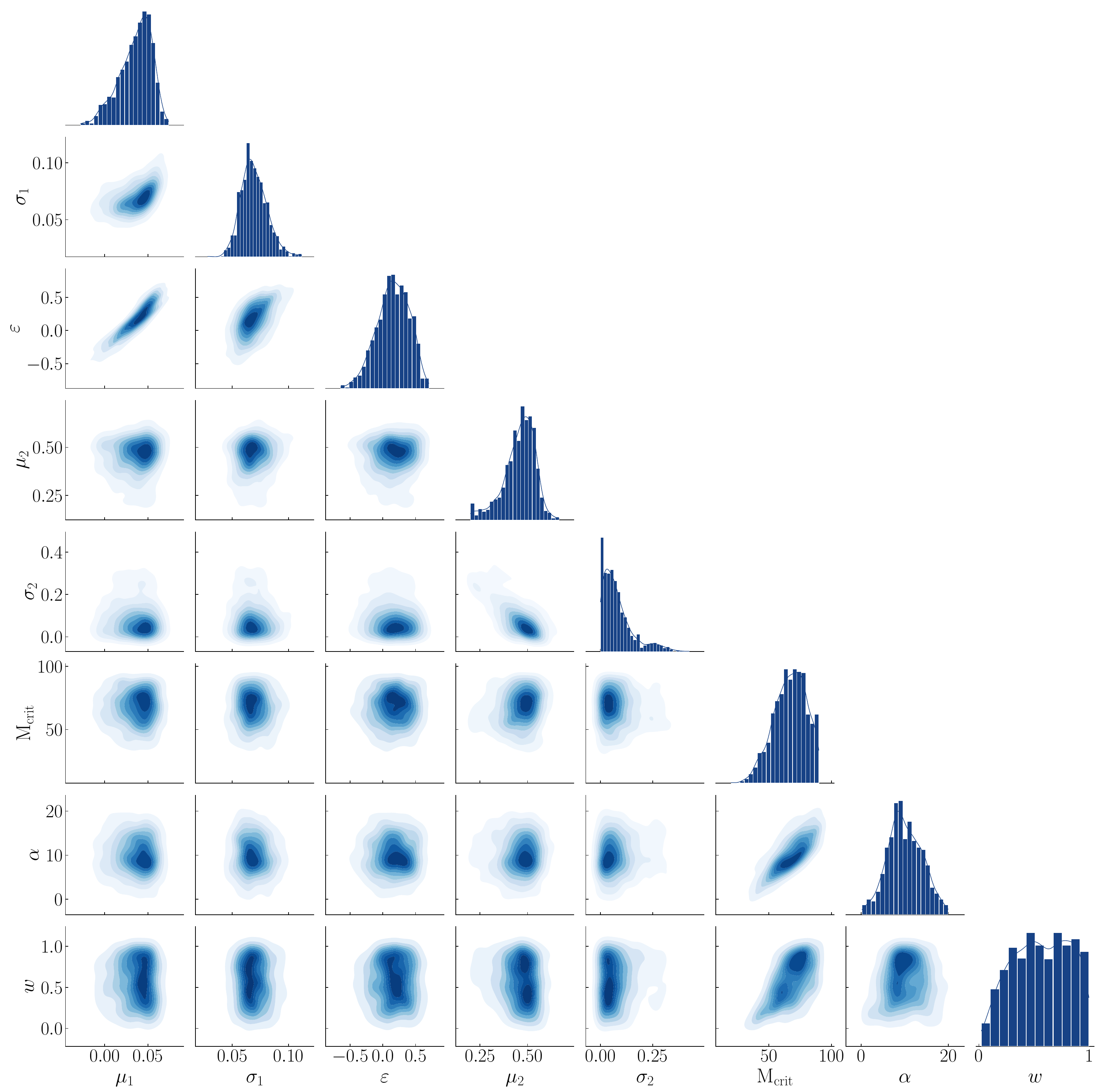}
    \caption{Joint posterior distribution for the parametrised, mass-dependent effective spin distribution.}
    \label{fig:cornerplot}
\end{figure*}

In this Appendix, we give the full functional form of the parametrised $\chieff$ distribution used in Section~\ref{sec:aligned} as well as the posterior distribution for its parameters. Following \citet{astrodist_gwtc4:2025}, we modelled the main peak of the effective spin distribution as a skewed Gaussian distribution
\begin{multline}
    \mathcal{SN}(\chieff|\mu_1,\sigma_1,\varepsilon)\\ \propto \begin{cases}
        (1-\varepsilon)\mathcal{N}\qty(\chieff|\mu_1,\sigma_1(1-\varepsilon))\,\mathrm{if}\,\chieff > 0\\
        (1+\varepsilon)\mathcal{N}\qty(\chieff|\mu_1,\sigma_1(1+\varepsilon))\,\mathrm{if}\,\chieff < 0
    \end{cases}
\end{multline}
where $\mathcal{N}(\chieff|\mu,\sigma)$ denotes a Gaussian distribution and $\varepsilon\in[-1,1]$ is the skewness parameter. Our model is a superposition of this skewed Gaussian and a second Gaussian distribution, weighted with a mass-dependent activation function
\begin{equation}\label{eq:activation}
    \mathcal{S}(\mathrm{M}_1|\mcrit,\alpha) = \frac{1}{1+e^{-(\mathrm{M}_1-\mcrit)/\alpha}}\,.
\end{equation}
Overall, the parametrised effective spin distribution reads:
\begin{multline}
    p(\chieff|\mathrm{M}_1) \propto \qty(1-w\,\mathcal{S}(\mathrm{M}_1|\mathrm{M_{crit}},\alpha)) \mathcal{SN}\qty(\chieff|\mu_1,\sigma_1,\varepsilon) \\ + w\,\mathcal{S}(\mathrm{M}_1|\mathrm{M_{crit}},\alpha) \mathcal{N}\qty(\chieff|\mu_2,\sigma_2)\,.
\end{multline}
The joint posterior distribution for the parameters of the mass ratio, redshift, and effective spin is reported in Figure~\ref{fig:cornerplot}

\section{BBH mergers in AGN discs}\label{app:AGNs}

\begin{figure*}
    \centering
    \includegraphics[width=0.8\linewidth]{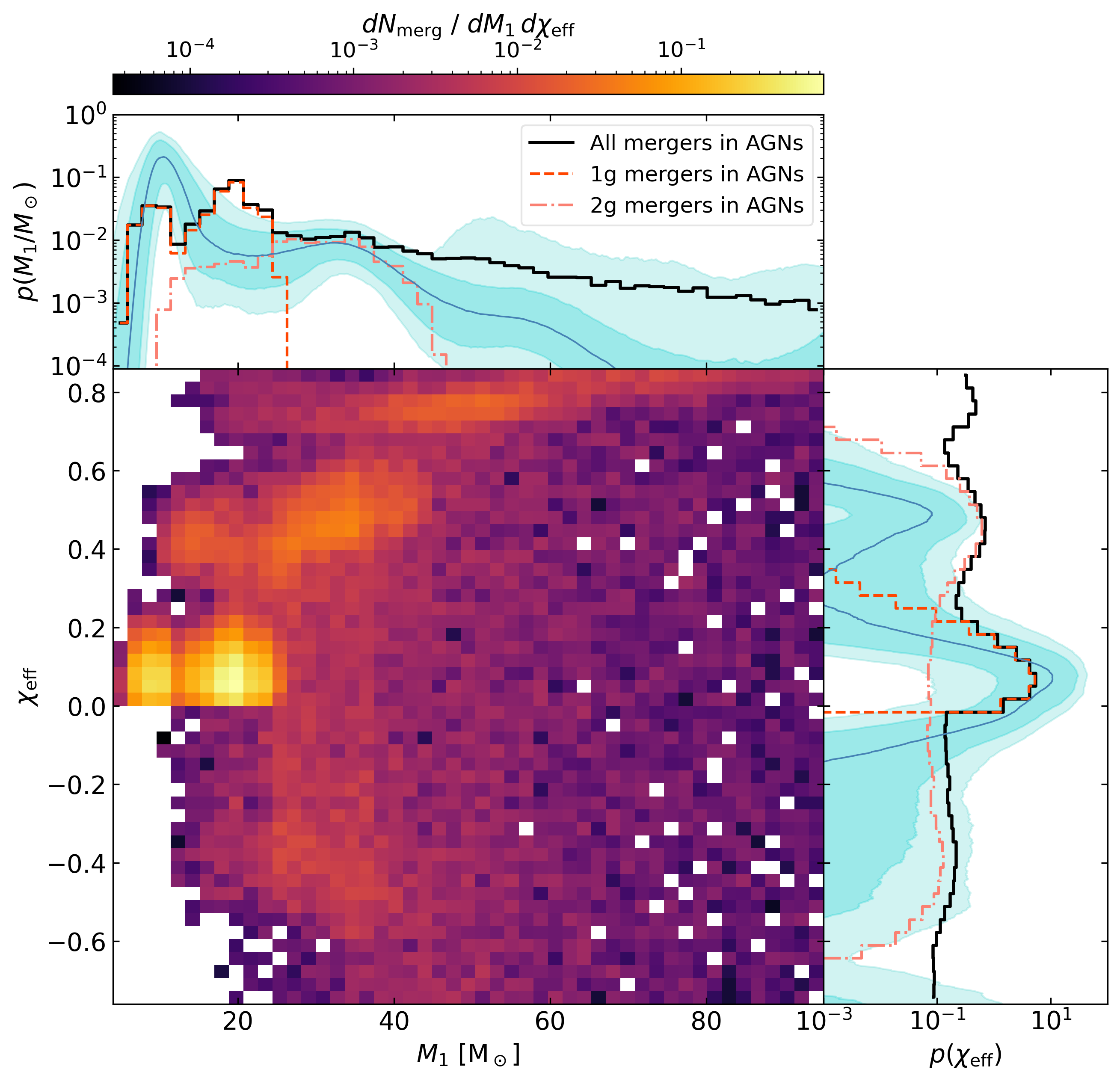}
    \caption{Distribution of primary mass, $\mathrm{M}_1$, and effective spin, $\chieff$, for the local AGN-disk merger population. The central panel shows the weighted two-dimensional merger distribution in the $(\mathrm{M}_1,\chieff)$ plane, normalized as $dN_{\rm merg}/ d\mathrm{M}_1  d\chieff$. The top and right panels show the corresponding marginalized distributions in $\mathrm{M}_1$ and $\chieff$, respectively. Black solid lines include all mergers in our AGN disc model, while the red dashed and dash-dotted lines show the contributions from first-generation and second-generation mergers, respectively. The cyan shaded regions and blue contours show the reference marginalized distribution obtained using (H)DPGMM (as in \autoref{fig:marginals}). The solid line marks the median distribution, whereas the shaded areas correspond to the 68\% and 90\% credible regions.}
    \label{fig:AGN_population}
\end{figure*}

Here we briefly summarize our semi-analytical AGN disk population model, presented in \citet{vaccaro:2024, vaccaro:2026}. 
We model the formation and hierarchical growth of BBHs in AGN disks with the \textsc{fastcluster} code\footnote{\textsc{fastcluster} is an open-source code available at \href{https://gitlab.com/micmap/fastcluster_open}{\texttt{https://gitlab.com/micmap/fastcluster\_open}}.} \citep{mapelli:2021, mapelli:2022, vaccaro:2024, torniamenti:2024}. 
The AGN disk structure is described with the steady-state model of \citet{SG} with viscosity $\alpha=0.01$ and SMBH sampled from \citet{greene:2007}. The initial stellar-origin BH masses are generated from \textsc{sevn} stellar-evolution models \citep{iorio:2023} assuming metallicity $Z=0.02$, while their spin magnitudes are drawn from a Gaussian distribution centered on zero, $p(\chi_1) \sim \mathcal{N}(0, \sigma_\chi)$, with $\sigma_\chi=0.05$ and truncated at $\chi\in[0,1]$.

The model follows stellar-mass BHs from their capture into the AGN disk to BBH formation and merger. 
BHs initially outside the disk plane are captured by hydrodynamical drag \citep{rowan:2025}. 
Once embedded, they migrate under gas torques as in \citet{vaccaro:2025}. 
When two BHs come within their mutual Hill radius, we determine whether they form a bound BBH using the pairing criterion of \citet{qian:2024}. If so, we assume that newly formed BBHs are prograde, with their orbital plane aligned with the AGN disc.

After formation, BBHs evolve under the combined effect of gas hardening, three-body encounters, and GW emission. 
Gas-driven evolution is modelled following \citet{ishibashi:2024}, while encounters with single BHs are treated by comparing the encounter timescale of \citet{leigh:2018} with the inspiral timescale. 
When an encounter occurs, its outcome is drawn from a grid of post-Newtonian three-body scattering experiments performed with \textsc{tsunami} \citep{tsunami, trani:2024}. 
At small separations, GW emission dominates and the binary is evolved to merger using the orbit-averaged equations of \citet{peters:1964}.

For each merger, we compute the remnant mass, spin, and recoil kick \citep{jimenez,maggiore_GW}. 
If the remnant remains bound to the AGN disk, it can undergo subsequent capture, migration, pairing, and further merger episodes. 
This recursive treatment enables the formation of hierarchical merger chains, which continue until the AGN lifetime is exceeded, the remnant is ejected by recoil, or the local supply of BHs is exhausted.

In \autoref{fig:AGN_population}, we compare the synthetic AGN-disk merger population with the reconstructed BBH population inferred from GWTC-5.0. The AGN model naturally populates the region of the $(\mathrm{M}_1,\chieff)$ plane associated with massive, positively aligned systems. In particular, hierarchical mergers in the disc produce a broad tail toward $\mathrm{M}_1 \gtrsim 40\,\msun$ and $\chieff \gtrsim 0.2$, overlapping with the high-mass aligned feature identified in the observed population. This qualitative agreement supports the idea that AGN discs can contribute to the formation of the aligned massive BBH subpopulation discussed in \autoref{sec:astro_implications}.

The agreement, however, is only partial. The AGN channel, as modelled here, fails to reproduce the primary mass peak at $10\,\msun$, while approaching the upper edge of the $90\%$ credible region of the inferred distribution for $\mathrm{M}_1\simeq 20\, \msun$ and $\mathrm{M}_1\gtrsim40\, \msun$. Moreover, the AGN channel populates the region with extreme $\chieff$, presenting prominent tails at large negative effective spin, $\chieff\lesssim -0.2$, and very large positive effective spin, $\chieff\gtrsim 0.5$. The excess at positive $\chieff$ is expected because BBHs formed in the disc are initially assumed to be prograde with the disc. The negative-$\chieff$ tail is instead mainly generated by later dynamical processing, in particular three-body encounters, which can perturb the orbital orientation of the binary before merger. As a result, the AGN model overpopulates the high-$|\chieff|$ regions compared to the reconstructed observed distribution.

This overpopulation is closely connected to the efficiency of hierarchical growth in the model. The most extreme regions of the $(\mathrm{M}_1,\chieff)$ plane are preferentially populated by higher-generation merger remnants, whose masses and spin magnitudes have been built up through repeated mergers in the disc. In particular, the ``$2$g'' population, where the primary BH is itself the remnant of a previous merger, already produces a visible peak at relatively large $\chieff\simeq 0.4$. Reducing the production or retention of higher-generation remnants, for instance through stronger recoil ejection, shorter AGN lifetimes, or less efficient pairing, would therefore mitigate the high-$|\chieff|$ tails and bring the AGN population closer to the inferred distribution. 

Overall, \autoref{fig:AGN_population} shows that AGN discs provide a plausible formation channel for the high-mass, aligned systems inferred in this work, but not for the full BBH population. If AGN discs were the dominant BBH formation channel, the predicted population would contain too many high-mass systems and too many binaries with large $|\chieff|$ compared to the observed distribution. A more consistent interpretation is therefore that AGN discs contribute only a subdominant fraction of the total BBH merger rate. In this case, the broad AGN mass-spin distribution can selectively populate the high-$M_1$, high-$\chieff$ region, while the bulk of the observed low-spin population is produced by other channels.

\end{appendix}
\end{document}